\documentclass{phb-proc4-auth}
\usepackage{graphicx,rotating,subfigure,amsmath,amsfonts,amssymb,delarray,cite,times}
\renewcommand{\vec}[1]{\boldsymbol #1}

\newcommand{\im}{\text{i}}

\begin{document}
\bibliographystyle{mc-elsart-num}
\begin{frontmatter}


\journal{SCES '04}


\title{Bose-Einstein condensation of magnons in TlCuCl$_3$}

%
%
%
%
%
%

\author{J. Sirker\corauthref{1}},  
\author{A. Wei{\ss}e} \and
\author{O.P. Sushkov}

%
 
\address{School of Physics, The University of New South Wales,
  Sydney 2052, Australia}

%
%
%
%


%
%
%
%

\corauth[1]{Corresponding author: sirker@phys.unsw.edu.au}


\begin{abstract}
 A quantitative study of the field-induced magnetic ordering in TlCuCl$_3$ in
  terms of a Bose-Einstein condensation (BEC) of magnons is presented. It is
  shown that the hitherto proposed simple BEC scenario is in quantitative and
  qualitative disagreement with experiment. It is further shown that even very
  small Dzyaloshinsky-Moriya interactions or a staggered $g$ tensor component
  of a certain type can change the BEC picture qualitatively. Such terms lead
  to a nonzero condensate density for all temperatures and a gapped
  quasiparticle spectrum. Including this type of interaction allows us to
  obtain good agreement with experimental data.
\end{abstract}

%
%



\end{frontmatter}

%
%
%
%
%

TlCuCl$_3$ is a spin-dimer system with an excitation gap $\Delta \approx
0.7$~meV in zero magnetic field. The dimers in this compound are formed by the
$S=1/2$ spins of the Cu$^{2+}$ ions and superexchange interactions are
mediated by the Cl$^-$ ions. The lowest excited state is a triplet of massive
bosons (magnons). The density $n$ of the boson mode is directly related to the
magnetization per site $m = g\mu_B n$. The magnon dispersion has been measured
by inelastic neutron scattering \cite{CavadiniHeigold} revealing a
considerable dispersion in all spatial directions. TlCuCl$_3$ therefore has to
be considered as a three-dimensional interacting dimer system.

\begin{figure}[!htp]
\begin{center}
\includegraphics*[width=0.7\columnwidth]{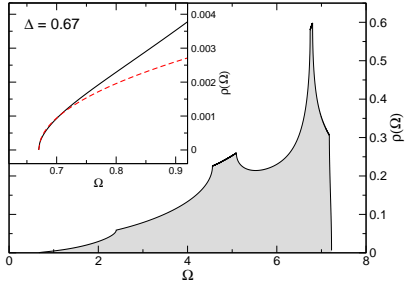}
\end{center}
\caption{Density of states $\rho(\Omega)$. The inset
  shows $\rho(\Omega)$ near the gap and a fit (dashed line)
  $\rho(\Omega)\sim\sqrt{\Omega-\Delta}$ yielding an effective mass
  $m_{\text{eff}}\sim0.2$ meV$^{-1}$.}
\label{fig1}
\end{figure}

A magnetic field $H$ causes a Zeeman splitting of the triplet with the lowest
mode crossing the singlet state at a critical field $H_c = \Delta/g\mu_B\sim
5.6$ T. The ground state for $H>H_c$ becomes a BEC of this low-energy boson.
The condensation of magnons corresponds to a phase transition into a state
with long-range antiferromagnetic order perpendicular to the applied field.
In Ref.~\cite{NikuniOshikawa} it has been shown that this transition and the
overall shape of the magnetization curves as a function of temperature are in
rough qualitative agreement with the BEC picture. However, we find that the
simple magnon dispersion $\Delta+\vec{k}^2/2m$ used in \cite{NikuniOshikawa}
is only justified for temperatures $T\ll 1$ K (see inset of Fig.~\ref{fig1})
well below the experimental temperature range. Therefore the universal
power-law dependence between critical density and temperature $n_c\sim
T_c^\alpha$ with $\alpha =3/2$ expected within the BEC picture will only be
observable at extremely low $T$ thus offering a simple explanation for the
deviations observed in \cite{NikuniOshikawa}. Following
Ref.~\cite{MatsumotoNormandPRL} we have therefore first calculated the real
triplet dispersion using a bond operator formalism \cite{SirkerWeisse}. The
starting point is the strong coupling ground state $|s\rangle$ where each
dimer at site $i$ is in singlet configuration $|i,s\rangle$.
One then expands the Hamiltonian in terms of operators $t^\dagger_{i\alpha}$
which create local triplet excitations with spiral index $\alpha$. In first
order a Hamiltonian bilinear in the operators $t^{(\dagger)}$ is derived. Only
the most important exchange pathways between the dimers are considered so that
the limited exchange parameter set can be determined by a fit to the measured
dispersion \cite{CavadiniHeigold}. In agreement with
\cite{MatsumotoNormandPRL} we find that terms quartic in
the triplet operators derived in next order yield only small corrections. The
most important renormalization of the triplet dispersion is due to the
hard-core constraint for the triplets, which can be taken into account by
introducing an infinite on-site potential
\begin{equation}
\label{eq1}
\mathcal{H}_U = U\sum_{i,\alpha,\beta}
t^\dagger_{i\alpha}t^\dagger_{i\beta}t_{i\alpha}t_{i\beta}, \quad
U\rightarrow\infty\; .
\end{equation}
The corresponding two-particle scattering vertex $v(\vec{k},\omega)$ can be
calculated exactly in the dilute limit by a summation of ladder diagrams
\cite{KotovSushkov}. Doing so we find a considerable renormalization of the
exchange parameters compared to the mean-field treatment in
\cite{MatsumotoNormandPRL}. For the scattering vertex at
the band minimum $\vec{q}_0$ we find $v_0=v(\vec{q}_0,0)\approx 10$ meV
\cite{SirkerWeisse}. The density of states $\rho(\Omega)$ calculated with the
renormalized dispersion $\Omega_{\vec{k}}$ is shown in Fig.~\ref{fig1}.

\begin{figure}[!htp]
\includegraphics*[width=0.99\columnwidth]{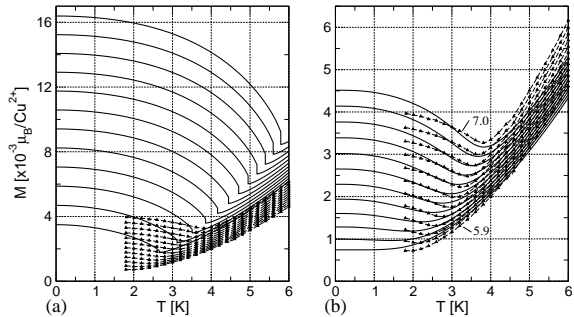}
\caption{Experimental $M(T)$ (symbols) for $H=5.9,6.0,\cdots,7$~T taken from
  Ref.~\cite{NikuniOshikawa}. (a) The theoretical $M(T)$ (solid lines) with
  $\Delta = 0.67$~meV and $v_0 = 9.8$~meV. (b) Interaction (\ref{eq3})
  included with $\gamma=10^{-3}$~meV, $\Delta = 0.72$~meV and $v_0 = 27$~meV.}
\label{fig2}
\end{figure}

For magnetic fields $H\gtrsim H_c$ and temperatures $T<\Delta$ it is
sufficient to take only the lowest triplet mode into account. The Hamiltonian
is then given by
\begin{equation}
\label{eq2}  
\mathcal{H} = \sum_{\vec{k}} \left(\epsilon_{\vec{k}}-\mu_0\right)t_{\vec{k}}^\dagger t_{\vec{k}}
+\frac{v_0}{2}\sum_{\vec{k},\vec{k}',\vec{q}} t_{\vec{k}+\vec{q}}^\dagger t_{\vec{k}'-\vec{q}}^\dagger
t_{\vec{k}} t_{\vec{k}'} 
\end{equation}
where $\epsilon_{\vec{k}}\equiv\Omega_{\vec{k}}-\Delta$ and $\mu_0 = g\mu_B
(H-H_c)$. In TlCuCl$_3$ typical magnon densities in magnetic fields $H\sim
6-7$ T are of the order $n\sim 10^{-3}$ \cite{NikuniOshikawa}. This allows us
to calculate the thermodynamic properties systematically in an expansion in
$n$ (gas approximation). Practically it turns out to be sufficient to treat
the interaction in (\ref{eq2}) in the one-loop approximation. After
diagonalizing the resulting bilinear Hamiltonian we can directly calculate the
Boson density (magnetization) as a function of temperature \cite{SirkerWeisse}
using the density of states from Fig.~\ref{fig1}. The result is shown in
Fig.~\ref{fig2}a.
Compared to experiment the theoretically calculated magnetizations are too
large and the $T_c$'s too high although $n_c\sim T_c^{2.1}$ yields a good fit
with the exponent being close to the observed one.

Finally, we want to discuss reasons for this shortcoming. First, apart from a
small temperature interval $|T-T_c|\lesssim n^{1/3}aT_c$ with
$n^{1/3}a\lesssim 0.1$ the one-loop approximation is reliable and higher order
corrections are small. Therefore there must be some physics we missed so far.
An interaction which we ignored and which is known to be important in
TlCuCl$_3$ is spin-phonon coupling \cite{ChoiGuentherodt}. Its main effect
will be a reduction of the bare magnon bandwidth and therefore an effective
increase of $v_0$. With $v_0\sim 25$ meV we can indeed obtain good agreement
with experimental data for $T\geq T_c$, however, the calculated magnetizations
at low temperatures are still 50\% too large \cite{SirkerWeisse}. In recent
ESR measurements \cite{GlazkovSmirnov} a direct singlet-triplet transition has
been observed which suggests that a perturbation of the form
\begin{equation}
\label{eq3}  
\mathcal{H}' = \im\gamma\left(t_{\vec{q_0}} - t^\dagger_{\vec{q_0}}\right) 
\end{equation}
might be present mixing the singlet and the triplet state. Here $\gamma$ is a
small parameter. Such a perturbation is equivalent to a staggered magnetic
field perpendicular to the applied field. It can result from a staggered $g$
tensor or Dzyaloshinsky-Moriya interactions which are both induced by
spin-orbit coupling. With this term included there is no longer a sharp phase
transition from a condensed into a non-condensed phase but instead the number
of condensed magnons decreases exponentially with temperature.  Furthermore
there is no longer a gapless mode usually characteristic for the condensed
phase \cite{SirkerWeisse}. In TlCuCl$_3$ the interaction (\ref{eq3}) requires
breaking of the inversion symmetry within a dimer. This might be due to small
static distortions as discussed in Ref.~\cite{SirkerWeisse}. We therefore
expect $\gamma$ to be quite small. It turns out that it is indeed possible to
obtain excellent agreement with experiment with $\gamma=10^{-3}$ meV as shown
in Fig.~\ref{fig2}b.

%
%
%
%

%
%
%
%


\end{document}